\documentclass[preprint,prb]{revtex4}

\usepackage{amsmath}
\newcommand{{\bff}}{\mathbf{f}}

\newcommand{\bfF}{\mathbf{F}}

\newcommand{\bfa}{\mathbf{a}}
\newcommand{{\bfg}}{\mathbf{g}}
\newcommand{{\bfv}}{\mathbf{v}}
\newcommand{{\bfA}}{\mathbf{A}}
\newcommand{{\bfV}}{\mathbf{V}}

\def\v#1{{\bf#1}}

\begin{document}

\title{Reply to ``Comment(s) on `Preacceleration without radiation: The non-existence of
preradiation phenomenon,' " by J. D. Jackson [Am. J. Phys. {\bf 75}, 844-845 (2007)] and V. Hnizdo [Am. J. Phys. {\bf 75}, 845-846 (2007)]}
\author{Jos\'e A. Heras}
\email{herasgomez@gmail.com}
\affiliation{Departamento de F\'isica y Matem\'aticas, Universidad Iberoamericana,
Prolongaci\'on Paseo de la Reforma 880, M\'exico D. F. 01210, M\'exico}
%\pacs{03.50.De, 03.50.Kk, 41.20.Cv, 41.20.Gz, 41.20.Jb}

\maketitle
Although I disagree with the comments of Jackson\cite{1} and Hnizdo\cite{2} on my recent paper\cite{3} on preacceleration without radiation, their objections are worth considering because they help to clarify certain important points not emphasized in my original paper.
 
Jackson\cite{1} states that ``It is unfortunate that Heras has chosen to discuss in detail the superseded Lorentz-Abraham-Dirac equation." I believe that not everybody would agree that the Lorentz-Abraham-Dirac equation is superseded. In my paper I discussed only the nonrelativistic limit of the Lorentz-Abraham-Dirac equation, that is, the Abraham-Lorentz equation: 
\begin{equation}
m\bfa=\v F_{\rm ext} + m\tau\dot{\bfa},
\end{equation}
which can hardly be considered superseded because it constitutes a formal result of classical electrodynamics. Equation~(1) can be derived by at least three independent methods: from energy conservation,\cite{4} from the self-force on an extended charge in the limit that it shrinks to a point,\cite{5} and from a general equation of motion obtained by assuming the point-electron model.\cite{6} It is true that it is difficult to understand how Eq.~(1) can predict acausal preacceleration. But it is also true that a judicious use of Eq.~(1) allows us to explore the effects of radiation on the motion of 
the point electron. 

Jackson\cite{1} also endorses the idea that the correct causal classical equation of motion for a quasi-particle in the nonrelativistic limit is given by 
\begin{equation}
\label{this}
m\bfa =\v F_{\rm ext} + \tau\dot{\v F}_{\rm ext}.
\end{equation}
He attributes Eq.~\eqref{this} to Landau and Lifshitz.\cite{7} Note that Eq.~(2) does not explicitly appear in any edition of their text. An approximate derivation of Eq.~(2) (which corresponds to an electron with structure)
from Eq.~(1) (which corresponds to a point electron)
has been suggested by Jackson.\cite{4} Does the electron acquire structure by the approximation process? In addition, Refs.~\onlinecite{2,3,4} in Ref.~\onlinecite{1} start with the ``superseded" Lorentz-Abraham-Dirac equation to obtain Eq.~(2). 
Fortunately, a cleaner understanding of Eqs.~(1) and (2) has been given by Ford and O'Connell\cite{6} who derived a family of solutions of a general equation of motion, each member of the family corresponding to an electron with a specified structure.
In particular, they showed exactly that Eq.~(1) corresponds to a point electron and that it violates causality. They also showed that causality demands that the electron should have a structure and that Eq.~(2) results from choosing the minimum structure allowed, which corresponds to an electron with a radius of the order of the classical radius of the electron. Thus, it is not necessary to make approximations to Eq.~(1) to derive Eq.~(2). For this reason, I suggest that Eq.~(2) be called the Ford-O'Connell equation.\cite{7} A reasonable conclusion is that the Ford-O'Connell equation applies to an electron with structure and the Abraham-Lorentz equation to a point electron. 

Jackson\cite{1} also strongly disagrees with my conclusion that for well-behaved external forces, the acausal radiation associated with the preacceleration (the preradiation) does not exist. He concludes that ``\ldots\ the preacceleration necessary to avoid runaway solutions produces acausal ``preradiation," contrary to Heras's claims." Let me first recall that the Larmor formula 
$ P_L=m\tau a^2$ allows us to calculate the total energy radiated in the interval $(t_1,t_2)$ by means of the expression
\begin{equation}
W_R(t_1,t_2)= m\tau\!\int_{t_1}^{t_2}\!a^2\,dt. 
\end{equation}
In Ref.~\onlinecite{3}, I derived Eq.~(8) for the nonrunaway solution of Eq.~(1) for well-behaved forces [given by Eq.~(7) of Ref.~3] acting during the time interval $(0,T)$. This nonrunaway solution is defined in two time intervals: the preacceleration interval $(-\infty,0)$ and the interaction interval $(0,T).$ I calculated the energies radiated in both intervals by applying Eq.~(3). By adding these two energies, I obtained the total radiated energy [Eq.~(8) of Ref.~3]
\begin{equation}
W_R(-\infty,\infty)= \frac{m\tau^2a(0)^2}{2} +
m\tau\!\int_0^T [\bfg(T)-\bfg(t)]^2 e^{2t/\tau}\,dt.
\end{equation}
The first term represents the acausal preradiation in the preacceleration interval, and the second term represents the radiation in the interaction interval. Both terms seem to be independent at first sight. Thus, Jackson's criticism seems to be correct. But I can use Eq.~(1) in Eq.~(3) to obtain the equivalent expression for the total energy [Eq.~(33) of Ref.~3]\cite{8}
\begin{equation}
W_R(t_1,t_2)= \tau\!\int_{t_1}^{t_2}\!\bfa\cdot\bfF_{\rm ext}\, dt +\frac{m\tau^2}{2}[a(t_2)^2- a(t_1)^2].
\end{equation}
I can now apply Eq.~(5) to the nonrunaway solution [Eq.~(8) of Ref.~3] to calculate the energies in the intervals $(-\infty,0)$ and $(0,T)$. The addition of these two energies again yields the total radiated energy,
\begin{align}
W_R(-\infty,\infty)\!= \frac{m\tau^2a(0)^2}{2}\!+
\!\tau\!\int_0^T[\bfg(T)\!-\!\bfg(t)] e^{t/\tau}\cdot\bff(t)dt-\frac{m\tau^2a(0)^2}{2}. 
\end{align}
The first term represents the acausal preradiation in the preacceleration interval, and the second and third terms represent the radiation in the interaction interval. The energy radiated in the preacceleration interval is not independent of the energy radiated in the interaction interval (because both energies have a common term). What is more important is that in Eq.~(6) we see that the preradiative term is identically canceled by the last term arising in the interaction interval. Thus, the total radiative energy is given by 
\begin{equation}
W_R(-\infty,\infty)=\tau\!\int_0^T[\bfg(T)-\bfg(t)] e^{t/\tau}\cdot\bff(t)dt.
\end{equation}
Therefore, the term representing the acausal preradiation does not exist in the final expression for the observed total radiated energy when calculated by means of Eq.~(5). If Jackson claims that Eq.~(4) supports the existence of preradiation, then I can equally claim that Eq.~(7) supports the nonexistence of preradiation. At first sight, both claims seem to be equally valid 
because Eqs.~(4) and (7) are equivalent. However, closer examination shows that Jackson's conclusion is misleading. The terms in Eq.~(4) are not independent as can be seen from Eq.~(6) and therefore separate interpretations of them are meaningless. In other words, the existence of the preradiative term should not 
be proclaimed without previously considering the radiation in the interaction interval. Equation~(6) shows that the acausal preradiative term is always a spurious quantity due to the separation of the nonrunaway solution into two time intervals. If this solution were divided into three time intervals, we would have two spurious terms (one of them not necessarily preradiative). The cancellation of spurious terms is a consequence of the continuity of the nonrunaway solution of Eq.~(1) for the case of well-behaved external forces (see Problem~11.19 of Ref.~5).

Jackson\cite{1} has proposed a thought experiment ``to show definitively" that Eq.~(1) implies radiation associated with the preacceleration. He imagined an experiment that according to him ``\ldots\ corresponds to the first term of Eq.~(10) [Eq.~(4) of the present reply]." In other words, he aims to theoretically measure the acausal preradiation represented by the first term of Eq.~(4). However, Jackson never considered in his thought experiment the second term of Eq.~(4). He apparently assumed incorrectly that both terms in Eq.~(4) are independent. But Eq.~(6) shows that both terms are not independent, and thus no physical interpretation should be given to either of the terms without considering the other one. By taking into account both terms we conclude that the preacceleration necessary to avoid runaway solutions does not produce acausal preradiation, contrary to Jackson's claim.

On the other hand, Hnizdo\cite{2} claims that Eq.~(1) implies the instantaneous 
power balance [Eq.~(1) in Ref.~2]
\begin{equation}
\v F_{\rm ext}\cdot\v v=\frac{d}{d t}\Big(\frac 12 mv^2 \Big)+m\tau\dot{\v v}^2-\frac{d}{d t}( m\tau\v v\cdot\dot{\v v}),
\end{equation}
and therefore all solutions of Eq.~(1) must satisfy Eq.~(8), including those that ``feature preacceleration." He argues that Eq.~(8) ``unambiguously demands that even a preaccelerating charge radiates according to the Larmor radiation-rate formula." 

In Ref.~\onlinecite{4} we learned that Eq.~(1) ``\ldots\ can be considered as an equation that includes in some approximate and time-averaged way the reactive effects of the emission of radiation." This statement is consequence of the fact that the radiative term $m\tau\dot{\v v}^2$ in Eq.~(1) was derived from an approximate and time-averaged process.\cite{4} This radiative term appears also in Eq.~(8) and therefore this equation is meaningful only when integrated over a time interval. We emphasize that the connection between radiation and radiation reaction is subtle and does
not hold instantaneously, but only in an approximate and time-averaged sense. Therefore, from the instantaneous Eq.~(8) we cannot directly conclude that a preaccelerating charge radiates as Hnizdo claims. 

Hnizdo\cite{2} argues that my result in Eq.~(7) expressing that ``the total radiated energy as an integral over the time interval in which the external force is nonzero does not necessarily mean that this energy is radiated only in that interval." To support his claim, Hnizdo integrates Eq.~(8) over all time and finds that the total radiated energy for the nonrunaway solutions can be written as [Eq.~(2) in Ref.~2]
\begin{equation}
W_R(-\infty,\infty)=\!\int_0^T \v F_{\rm ext}\cdot\v v dt- \frac 12 m[\v v(\infty)^2-\v v(-\infty)^2].
\end{equation}
He states that Eq.~(9) ``\ldots\ gives no indication of the time interval in which any part of the energy $W_R(-\infty,\infty)$ is radiated."
But Eq.~(9) can be transformed into Eq.~(7) which indicates that the energy is radiated only in the interval $(0,T)$. 
Consider the velocity given by Eq.~(38) of Ref.~3 for the interval $0\leq t\leq T$, that is, $\v v(t)=\bfV(t)-\bfV(0)+\tau [\bfg(T)-\bfg(t)] e^{t/\tau}$, where $\bfV(t)$ is the velocity when the radiation reaction is not considered. By substituting this velocity into Eq.~(9) we obtain
\begin{align}
W_R(-\infty,\infty)\!=\!\int_0^T [\bfV(t)\!-\!\bfV(0)]\cdot \v F_{\rm ext}\,dt\! -\!\frac 12 m {\v v(\infty)}^2\!+\!\tau\int_0^T 
[\bfg(T)\!-\!\bfg(t)] e^{t/\tau}\cdot\v F_{\rm ext}\,dt.
\label{10}
\end{align}
Integration of the first term yields $m[\v v(T)]^2/2$, which cancels the second term of Eq.~\eqref{10} because of 
$\v v(T)=\v v(\infty)$, and thus Eq.~(10) reduces to Eq.~(7). There are different expressions for the total radiated power associated with the nonrunaway solutions of Eq.~(1) for well-behaved forces, some of which eventually involve spurious terms like the expression in Eq.~(4). We can show that any of these expressions can be transformed into Eq.~(7) which in turn gives the total radiated power without any spurious terms. In other words, after eliminating all spurious terms, the total radiated energy is emitted only in the interval in which the force acts and this result is properly represented by Eq.~(7), contrary to Hnizdo's claim.\cite{9}

Let me make a final and speculative comment. It appears that there is something wrong in Eq.~(3) when applied to the preacceleration interval alone because we obtain an acausal radiation. Since Eq.~(3) is defined by the Larmor formula $ P_L=m\tau a^2$, it follows that this well-known formula could not be always a correct measure of the instantaneous power radiated.\cite{10} Moreover, the Larmor formula may be questioned because it was derived by neglecting the radiation reaction.\cite{11}  Therefore, in the context of radiation reaction, a modification of the Larmor formula seems to be desirable and necessary. Let me suggest such a modification. I propose the following formula for the instantaneous power radiated:
$P_H= P_L -(\tau/2)dP_L/dt,$ or equivalently,
\begin{equation}
P_H= m\tau a^2 -\frac{d}{dt}\left(\frac{m\tau^2a^2}{2}\right). 
\end{equation}
We can show that the power $P_H$ is equivalent to the power $P_L$ in the sense that using energy conservation, $P_H$ also leads to the Abraham-Lorentz force $\v F_{\rm rad}=m\tau\dot{\v a}$. In fact, with the power  $P_H$ expressed as $P_H= -m\tau \v v\cdot\dot{\v a} +dS/dt-dT_R/dt$, where 
$S=m\tau\v v\cdot\v a$ is the Schott energy and $T_R= m\tau^2a^2/2$ is the radiative kinetic energy, energy conservation requires
\begin{equation}
\int_{t_1}^{t_2}\!\v F_{\rm rad}\cdot\v v\! =\! m\tau\!\int_{t_1}^{t_2}\!\dot{\v a}\cdot\v v\, dt \!-\!S(t_2)\!+\! S(t_1)\! +\!T_R(t_2)\!-\!T_R(t_2).
\end{equation}
If the motion is such that $S(t_2)=S(t_1)=0$ and $T_R(t_2)=T_R(t_1)=0$ then we can infer the Abraham-Lorentz force $\v F_{\rm rad}=m\tau\dot{\v a}$ from Eq.~(12). Using the Abraham-Lorentz equation, the power $P_H$ in Eq.~(11) can be expressed as 
\begin{equation}
P_H = \tau\bfa\cdot\bfF_{\rm ext}. 
\end{equation}
With this power the total energy radiated in the interval $(t_1,t_2)$ is given by\cite{12}  
\begin{equation}
W_{RH}(t_1,t_2)=\tau\!\int_{t_1}^{t_2}\!\bfa\cdot\bfF_{\rm ext}\,dt. 
\end{equation}
If we apply Eq.~(14) with $t_1=-\infty$ and $t_2=\infty$ to the acceleration given by Eq.~(8) of Ref.~3, which implies $S(-\infty)=S(\infty)=0$ and $T_R(-\infty)=T_R(\infty)=0$, then we obtain Eq.~(7) of this reply. Instead of $ P_L$, we could reasonably consider $P_H$ as a more appropriate formula for the instantaneous radiated power involving radiation reaction. The powers $P_H$ and $P_L$ are equivalent in the sense that both of them lead to the same total radiated energy associated with the acceleration in Eq.~(8) of Ref.~3. But $P_L$ implies an acausal spurious term when applied to the preacceleration interval, and $P_H$ does not imply this term. Therefore, instead of stating that ``according to the Larmor power $ P_L=m\tau a^2,$
an accelerated charge radiates," we could state that ``according to the power $P_H= \tau\bfa\cdot\bfF_{\rm ext},$ an accelerated charge radiates only when an external force acts."

\begin{acknowledgments}
I am grateful to an anonymous referee for his valuable comments on my reply and to Professor R.\ F.\ O'Connell for useful discussions on the Ford-O'Connell equation.
\end{acknowledgments}


\begin{thebibliography}{99}
\bibitem{8} To obtain Eq.~(5) from Eqs.~(3) and (1), I have performed an integration by parts. In his comment, Jackson claims that ``Integrations by parts are not physics." I am surprised by this statement considering that integration by parts is a well-recognized mathematical procedure which is used extensively in physics, in particular by Jackson who obtained the Abraham-Lorentz force: $\v F_{\rm rad}=m\tau\dot{\bfa}$ by performing an integration by parts (see Ref.~4).
\bibitem{10}
A referee has emphasized  that radiation is the stuff that eventually makes it out to infinity, and attributing that to a particular instant of emission, as the Larmor formula does, is always dangerous. In fact, our concept of radiation arises from considering the radiation fields that go like $1/R$, constructing from them the $1/R^2$ part of the Poynting vector, integrating over a large spherical surface and taking the limit as $R\to\infty.$ 
He also argues that all we really know is the total radiated power in $(-\infty, \infty)$, and how much of it we assign to the preacceleration regime $(-\infty, 0)$, and how much to the interval $(0,\infty)$, is on some level, a nonphysical question.
\bibitem{11}
The Larmor formula is derived from the radiation fields of a nonrelativistic point electron. The acceleration contained in these fields  
does not include an explicit contribution due to the radiation reaction force.
\bibitem{12} 
In Ref.~3 Hnizdo states that the power implied by Eq.~(2) is given by $\tau\v a\cdot\v F$, which seems to be the same power given by Eq.~(13). Nevertheless, they are different because the accelerations in the two expressions for the power are not the same.
\end{thebibliography}
\end{document}